\documentclass{article}
\usepackage{amsmath,graphicx}
\usepackage{hyperref}
\usepackage[textsize=scriptsize,obeyFinal,final]{todonotes}
\usepackage{INTERSPEECH2020}
\usepackage[T1]{fontenc}
\usepackage{CJKutf8} 

\newenvironment{Japanese}{
  \CJKfamily{min}%
  \CJKtilde
  \CJKnospace}{}

\title{SINGING SYNTHESIS: WITH A LITTLE HELP FROM MY ATTENTION.}

\name{Orazio Angelini \qquad Alexis Moinet \qquad Kayoko Yanagisawa \qquad Thomas Drugman}
\address{Amazon Research, Cambridge}
\email{\{aorazio, amoinet, yakayoko, drugman\}@amazon.com}

\newcommand{\qt}[1]{``#1''} 
\newcommand{\periphrasis}{AS2S}
\newcommand{\modelname}{UTACO}

\newcommand{\reffigure}[2][]{Figure#1 \ref{#2}}
\newcommand{\reffig}[2][]{Fig#1.\ref{#2}}

\newcommand{\refsec}[1]{Sec.\ref{#1}}

\newcommand{\diffblock}[1]{#1}

\begin{document}
\maketitle

\begin{abstract}
We present \modelname, a singing synthesis model based on an attention-based sequence-to-sequence mechanism and a vocoder based on dilated causal convolutions.
These two classes of models have significantly affected the field of text-to-speech, but have never been thoroughly applied to the task of singing synthesis.
\modelname\ demonstrates that attention can be successfully applied to the singing synthesis field and improves naturalness over the state of the art.
The system requires considerably less explicit modelling of voice features such as F0 patterns, vibratos, and note and phoneme durations, than previous models in the literature.
Despite this, it shows a strong improvement in naturalness with respect to previous neural singing synthesis models.
The model does not require any durations or pitch patterns as inputs, and learns to insert vibrato autonomously according to the musical context.
However, we observe that, by completely dispensing with any explicit duration modelling it becomes harder to obtain the fine control of timing needed to exactly match the tempo of a song.
\end{abstract}



\section{Introduction}
Research efforts on computer-aided singing synthesis systems date back to the late 1950s \cite{Cook1996}.
Historically, the working principles of singing synthesis systems have been based on parametric text-to-speech (TTS) or unit selection technology.
Notable recent examples are Sinsy \cite{Hono2018}, a statistical parametric singing synthesis system, and Vocaloid \cite{Kenmochi2007}, based on unit selection.

A recent development in the field is the introduction of deep neural networks (DNN) \cite{Gomez2018arXiv}.
The latest version of Sinsy adopts DNNs \cite{Nakamura2019arXiv}, and DNN sub-models exist to predict specific features of a singing voice, such as F0 \cite{Wada2018} or note transition and sustain patterns \cite{Hua2018arXiv}.
The introduction of the WaveNet architecture \cite{vandenoord2016} marked an increase in the importance of DNN techniques for TTS.
Singing synthesis followed, with the introduction of several DNN-based models. Examples include \cite{Blaauw2017arXiv,Blaauw2017,Bous2019arXiv}, which present variations on a model of a singing voice based on WaveNet\cite{vandenoord2016}, and \cite{Chandna2019} developing a WGAN architecture.
A common feature of all these singing synthesis models is the need to develop a number of separate specialised sub-models to predict specific voice features such as the F0 contour, the duration of individual phonemes, or the start time of notes (which, in natural singing voices, do not follow exactly the timing of the score \cite{Hono2018}).

A development in TTS technology that is relevant to our work has been the introduction of attention-based architectures \cite{Vaswani2017} such as Tacotron \cite{Wang2017arXiv,Shen2018} and Deep Voice \cite{Ping2017}, attention-based sequence-to-sequence (\periphrasis) models which predict spectrograms that are subsequently used to synthesise a waveform with a vocoder.
For our purposes, the most salient feature of \periphrasis\ architectures is that its only conditioning input is text (or a corresponding phoneme list), and not any additional model or piece of information.
Whereas previous models needed to be conditioned on several other pieces of context, for example F0, an \periphrasis\ autonomously learns an implicit model of all voice features that are not included in its inputs: e.g. intonation, stress and rhythm. This point is made explicitly in \cite{Wang2018arXiv,Skerry2018arXiv,Klimkov2019}, which try to learn an explicit representation for these features to be used in later conditioning. \cite{Valle2019} even uses it in singing, with an attention model that is fed pitch and duration externally.

In this paper, we consider the possibility that an \periphrasis\ architecture may be able to learn an implicit model of singing interpretation in a similar way to what it does for speech prosody. We train an \periphrasis\ architecture on singing data, and observe that it is capable of generalising to unseen musical scores. We find that \periphrasis\ architectures, conditioned only on the information available in a score, are capable of singing synthesis, and they can significantly improve naturalness with respect to the state of the art. We name our model \modelname, from the Japanese word \begin{CJK}{UTF8}{}\begin{Japanese}\qt{歌}\end{Japanese}\end{CJK}(\qt{uta}), meaning song, and the beginning of the word \qt{Tacotron}.

\section{System description}
\label{sec:system_description}
\begin{figure}[hbt]
\includegraphics[width=.47 \textwidth]{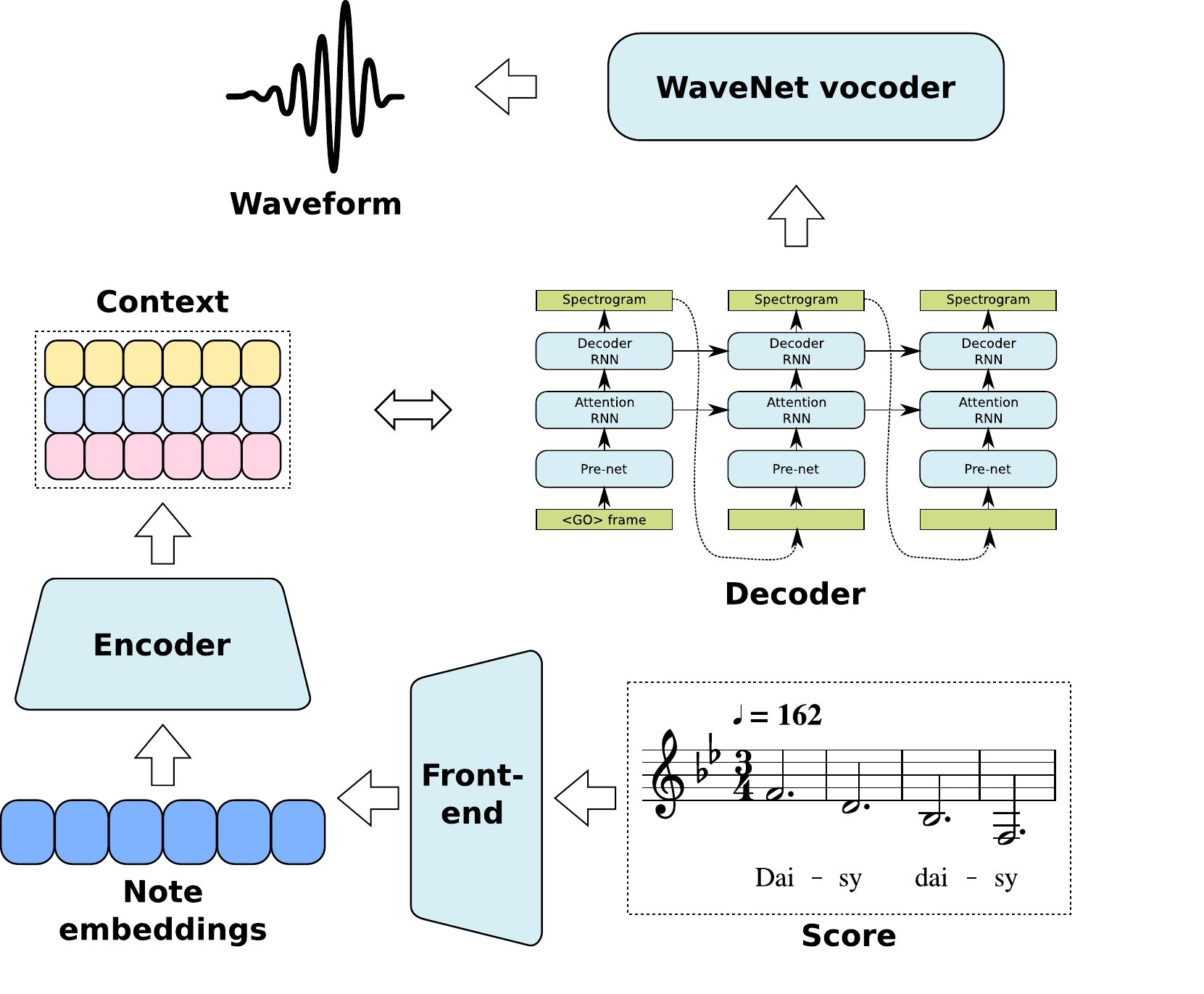}
\caption{Diagram of \modelname's arhitecture.} 
\label{fig:diagram}
\end{figure}

\modelname\ consists of three main parts (see \reffig{fig:diagram}). The first is a frontend that receives a score in MusicXML (MXL) format \cite{Good2006} as input, and outputs the note embeddings to be sent to an attention encoder.
The second is an \periphrasis\ architecture, based on \cite{Merritt2018,Latorre2018}, modified to accept the note embeddings, whose decoder produces mel-spectrograms.
The spectrograms are finally synthesized with a vocoder.

The frontend performs linguistic analysis on the score lyrics.
The phoneme sequence for the utterance is inferred from the lyrics text.
We allow for 3 possible vowel levels of stress (0,1,2).
The punctuation is ignored.
Then the frontend determines which phonemes correspond to each note of the score, using syllabification information specified in the MusicMXL file.
It also computes the expected duration in seconds of each note given its length, the tempo and time signature of the score.
It finally combines this information into embeddings that will be used to condition the \periphrasis\ model.

\begin{figure}[hbt]
\includegraphics[width=.49 \textwidth]{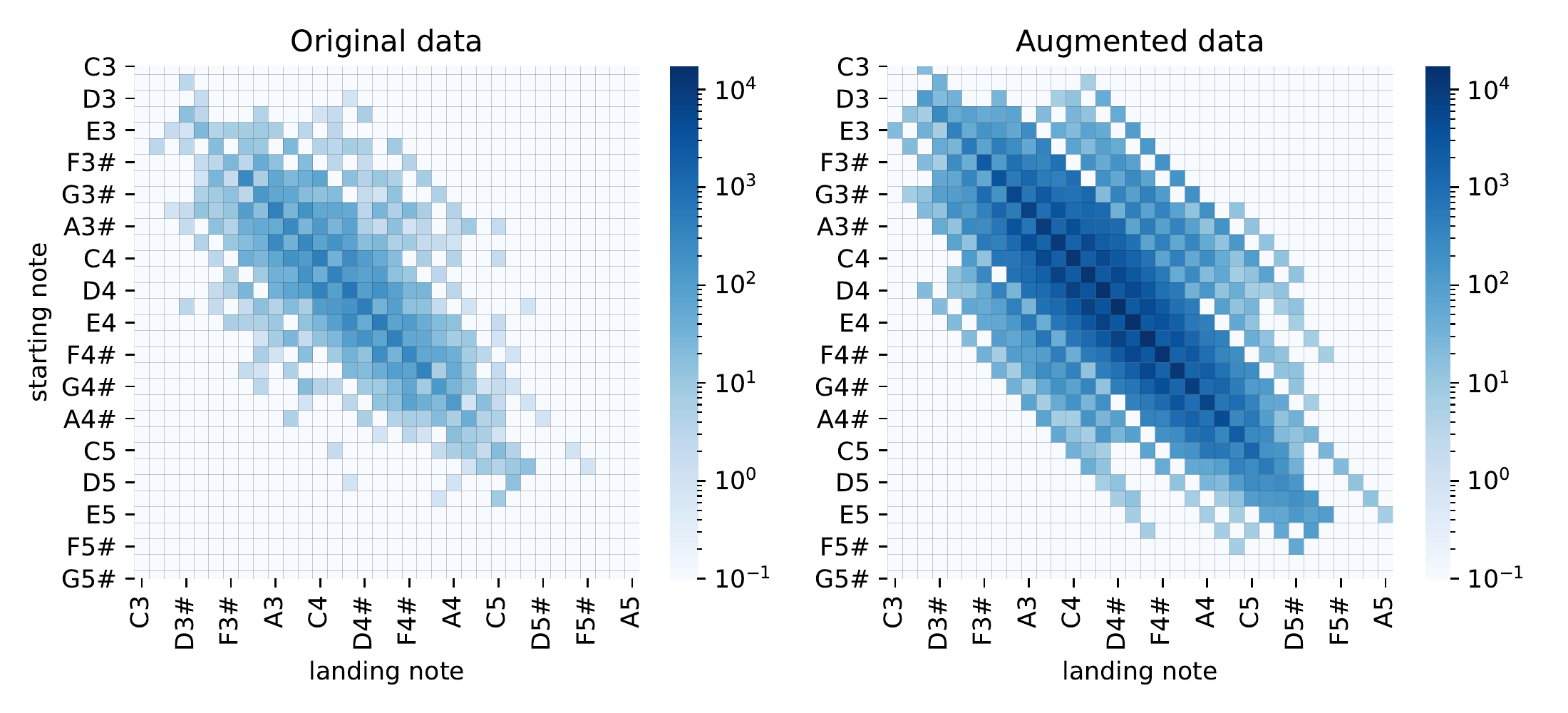}
\caption{Distribution of pitch changes in the dataset. Left (right) panel shows the distribution of the original (of augmented) data.}
\label{fig:pitch_shift}
\end{figure}

The modification applied to the TTS \periphrasis\ architecture concerns the conditioning inputs to the encoder.
Whereas a TTS \periphrasis\ generally takes as its only input a sequence of (one-hot encoded) tokens representing phoneme IDs for the utterance to be generated, our system uses phoneme-level note embeddings. These consist of 5 streams, all of equal length and concatenated:

\diffblock{\begin{enumerate}
\item The phoneme sequence for the song utterance to be generated, one-hot encoded. 84 tokens are available in this stream, including a start (\texttt{<s>}), and a word boundary (\texttt{<wb>}) one.
\item The octave sequence for the note to be sung on each phoneme, according to the score, one-hot encoded. For example, for the sequence of notes C4, D\#4, G3, the corresponding octave sequence would be (4, 4, 3). We allow for 4 values, which is the range found in our dataset.
\item The step in the chromatic scale (out of 12 possible ones) for the note to be sung on each phoneme, according to the score, e.g. G\#, or B-, one-hot encoded.
\item The duration in seconds of the note to be sung on each phoneme, represented as a floating point number (z-score normalised).
\item A position embedding computed as a ramp representing the advancement of the note for each phoneme that it contains, as a floating point number in the interval [0,1]. For example, if three phonemes have to be sung on a given note, the first phoneme will have 1.0 on this stream, the second 0.5, and the last 0.0.
\end{enumerate}}
\begin{figure}[hbt]
\includegraphics[width=.48 \textwidth]{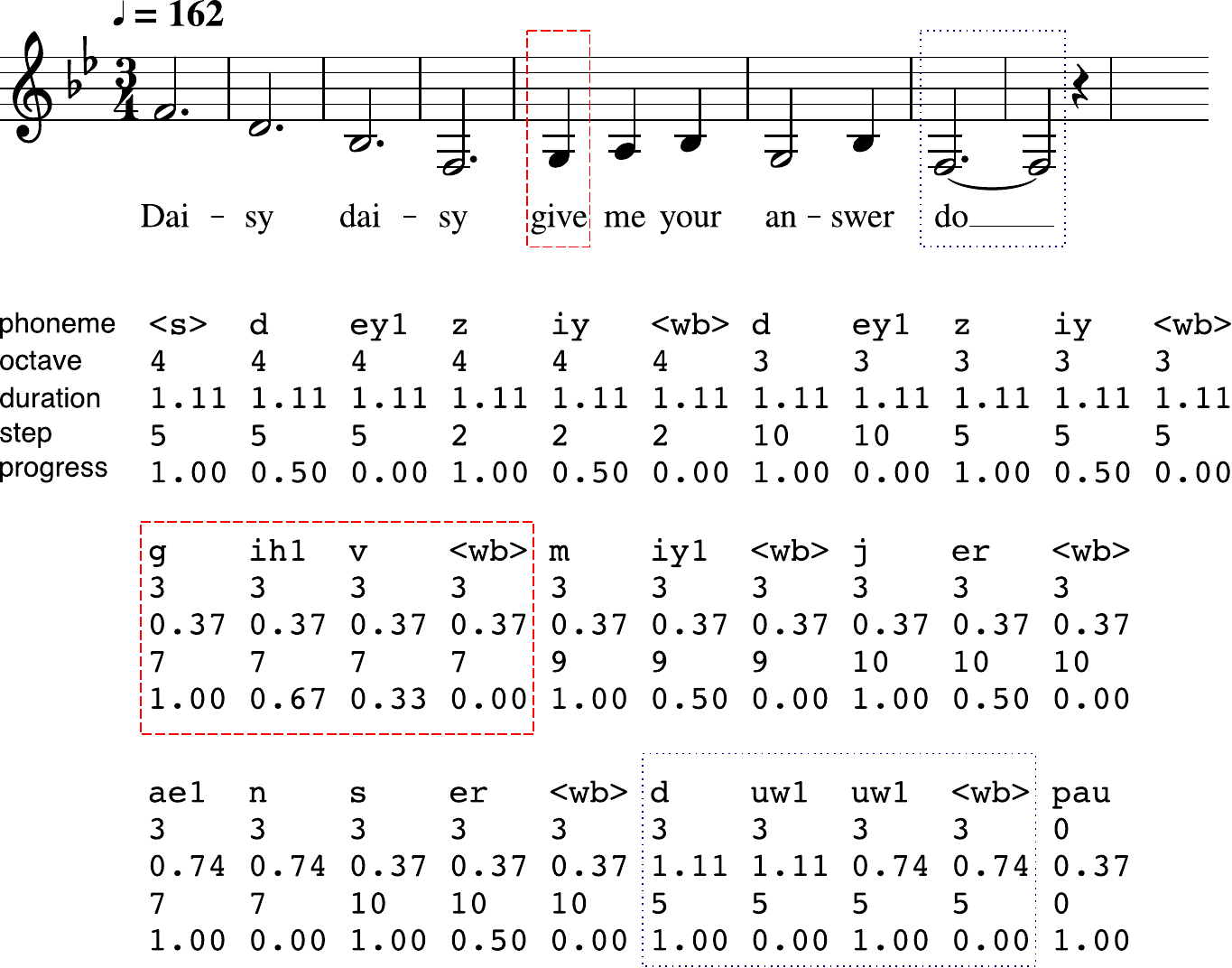}
\caption{An example of the embeddings detailed in \refsec{sec:system_description}. Dashed red highlights the example illustrated in \refsec{sec:system_description}, and dotted blue shows a tie disambiguated by stream 5.}
\label{fig:embeddings}
\end{figure}
Streams 2-4 are repeated for the length of the note. For example, if the word \qt{give} is to be sung on a G3 for 0.37 seconds, the tokens (\qt{\texttt{g}}, \qt{\texttt{ih1}}, \qt{\texttt{v}}, \qt{\texttt{<wb>}}) will be put in stream 1 (as in \reffig{fig:embeddings}), stream 2 will contain the octave (3, 3, 3, 3), the corresponding positions in stream 3 will contain (G, G, G, G), and those in stream 4 will contain the z-score normalised values corresponding to (0.37, 0.37, 0.37, 0.37).
Stream 5 will contain (1.0, 0.67, 0.33, 0.0).
If a phoneme has to be sung on several notes, the phoneme is repeated while the other streams change. For example, to sing the phoneme \qt{\texttt{uw1}} on two notes, G and E, each of duration 0.1 seconds, followed by one 0.2 second F, then stream 1 will contain (\qt{\texttt{uw1}}, \qt{\texttt{uw1}}, \qt{\texttt{uw1}}); while stream 3 will contain (G, E, F) and stream 4 (0.1, 0.1, 0.2).
The ramp also helps disambiguate ties, such as the one in \reffig{fig:embeddings}.
Rests are represented by a pause token, and an additional one-hot value for octave and step.
An example of the embeddings produced can be seen in \reffigure{fig:embeddings}.

Compared to analogous solutions, we avoided the use of embeddings in the style of \cite{huang2018music}, which are more useful in polyphonic music. Our inputs are less rich compared to \cite{Blaauw2017}, which will be discussed in \refsec{sec:discussion}.

The rest of the architecture is an \periphrasis\ model based on \cite{Latorre2018}. It was trained with the Adam optimisation algorithm \cite{Kingma2014} and a learning rate of 0.001 for $\sim$300K steps.
The network so organised is capable, after standard training on a large enough dataset, to produce spectrograms which are then synthesised by a vocoder, see \refsec{sec:experimental:dataset} and \ref{sec:experimental:vocoder} for more detail.

\section{Experimental protocol}
\subsection{Dataset}
\label{sec:experimental:dataset}
The dataset consists of 96 songs in US English, sung a cappella by a single female voice, for a total of 2 hours and 15 seconds of music.
It covers several genres, such as pop, blues rock, and some children's songs.
The songs have been autotuned to correct the performer's pitch errors.
Since the length of most songs is in the order of minutes, we split them into segments of $\sim$20-30 seconds, which correspond to $\sim$200 phonemes. This reduces the memory requirements at training time compared to processing whole songs -- due to the attention matrix size increasing with the square of the sequence length -- while keeping the batch size large enough, at 32.
In the vocoder training set we used an additional $\sim$40 hours of speech data by the same speaker performing the songs.

Given the small amount of data available, we reduced the test set size to the minimum possible, of about 5 minutes in total.
Most songs contain repetitions such as refrains or repeated pitch patterns.
Holding out whole songs would have been too costly in terms of training data, and yielding too little diversification in the test set.
In order to ensure a strict separation of train and test data, we compared all possible pairs of segments. If two segments included a sequence of 3 or more identical subsequent note pitches in the score, we excluded both from the test set.
Apart from the segments excluded because of repetitions, we selected the test set segments randomly.

\subsection{Data augmentation}
\label{sec:experimental:augmentation}
A common concern in the singing synthesis field consists of the small amount of data available to train models \cite{Blaauw2017}.
Song data is symmetric to two transformations: pitch shifting and tempo changes.
Following previous literature \cite{Blaauw2017,Ko2015,Nakamura2019arXiv}, we employed both. We applied the following transformations to each song in the dataset:
\diffblock{\begin{itemize}
 \item Shifting the pitch by [-1, 0, +1, +2, +3] semitones.
 \item Changing the original beats per minute (bpm) of the song to [.85, .90, .95, 1., 1.05, 1.10, 1.15] percent of the original one.
\end{itemize}}
There are 35 possible combinations of these two augmentation types, so the final amount of augmented data consisted of about 70 hours.
This makes more contexts available to the model during training.
The change can be visualised in \reffigure{fig:pitch_shift}.
We applied these transformations using an algorithm that preserves perceived vocal tract length. The maximum amount of change that can be applied before excessive degradation has been determined through informal listening tests by the authors.
Despite our attempts, the model without augmentations was unstable and produced mediocre quality when inferring on training data, and produced random phonemes and melody at inference on unseen inputs.
We believe that this is because the small number of contexts available in the unaugmented datasets
makes the model unable to extrapolate to unseen data.
Therefore, we did not include the model trained on unaugmented data in our experimental validation.

\subsection{Vocoder}
\label{sec:experimental:vocoder}
The vocoder used in this paper is an autoregressive WaveNet based on \cite{vandenoord2016}, conditioned only on (80-dimensional) mel-spectrograms as in \cite{Shen2018}.
The training data for the vocoder used in our test consisted of the whole training set (including augmentations) used for the \periphrasis\ model plus $\sim$40 hours of speech data by the same speaker.
We observed that the addition of speech data to the vocoder training set seems to increase the quality of the samples.

\subsection{Baseline}
\label{sec:baseline}
We compared \modelname\ to WGANSing \cite{Chandna2019}, the only recent state-of-the-art (SOTA) neural singing synthesis model for which a training protocol was released.
The system is inspired by Deep Convolutional Generative Adversarial Networks (DCGAN), with a Wasserstein-GAN loss, and uses the WORLD vocoder \cite{morise2016world}.
We compared our proposed system with WGANSing trained on the same (augmented) dataset as our model, using a slightly modified version of the protocol in the published repository \footnote{\url{https://github.com/MTG/WGANSing}, commit \texttt{pc2752}.}.
Specifically, the corpus used in \cite{Chandna2019} includes both a sung and spoken version of the same text for each utterance.
In the published protocol, both are used for training.
Since our dataset does not have the necessary spoken data, we only trained WGANSing on song data.
The effect of this change is discussed in \refsec{sec:baseline_consistency}.
Note that WGANSing needs to be fed external duration and continuous pitch, which we extracted from the original recordings.
This needs to be kept in mind for a fair comparison, since \modelname\ generates its own timing and pitch patterns.

\subsection{MUSHRA methodology}
\label{sec:experimental:mushra}
We set up MUSHRA tests \cite{Mushra2015} comparing various versions of each segment sung or synthesised in the same voice.
As recommended in \cite{Mushra2015}, the segments chosen for the test set were further split into chunks, each $\leq$10 seconds long. Cuts were made in naturally occurring pauses in the songs, trying to keep the segments as long as possible.
This generated typical lengths of $\sim$3-8 seconds.
Each segment was judged by US English native speakers, who were asked to \qt{rate the samples in terms of their naturalness}.
They rated the samples between 0 (representing \qt{Not at all natural}) and 100 (representing \qt{Completely natural}).
In all experiments, all possible combinations of pairwise 2-sided t-tests on the means of the scores for different systems yield a p-value $\ll$0.001.

\subsubsection{MUSHRA for baseline validation}
\label{sec:baseline_consistency}
\begin{figure}
\centering
\includegraphics[width=\textwidth / 2]{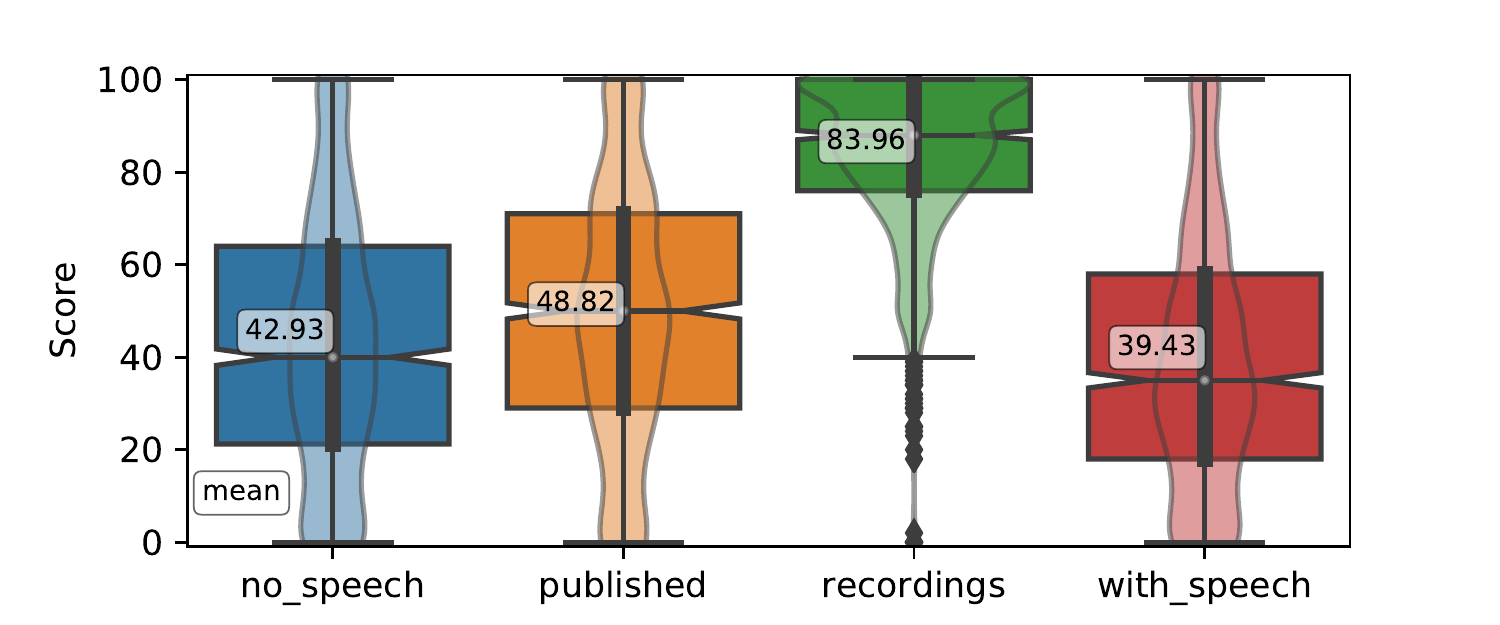}
\caption{Box and violin plot of MUSHRA for the baseline, trained on WGANSing's corpus.}
\label{fig:mushra_boxplot_baseline}
\end{figure}

We checked that training WGANSing without any spoken data does not alter the model performance and is therefore representative of the model's quality.
We trained WGANSing on the same corpus\footnote{\url{https://smcnus.comp.nus.edu.sg/nus-48e-sung-and-spoken-lyrics-corpus/}}
used by the authors in their publication~\cite{Chandna2019}.
We compared, in a separate MUSHRA test, their published audio clips\footnote{\url{https://pc2752.github.io/sing_synth_examples/} (last updated \texttt{03/02/2019})}
to the ones we produced.
To replicate our MUSHRA methodology, the audio clips have been cut into 51 segments, and each has been evaluated by 30 listeners.
We trained a model with speech data and another one without.
The results are shown in \reffig{fig:mushra_boxplot_baseline}.
Surprisingly, WGANSing with no speech in the training data (\texttt{no\textunderscore speech}) has a better score than the model with speech (\texttt{with\textunderscore speech}).
The difference in mean MUSHRA score between our best model (\texttt{no\textunderscore speech}) and the results published by the authors (\texttt{published}) is 5.89. The two score distributions for \texttt{no\textunderscore speech} and \texttt{published} are very similar, except for slightly fatter tails on the lower and higher end of the score spectrum, respectively.
Therefore, we conclude that training WGANSing on singing data alone does not deteriorate the model's overall performance.

\subsubsection{Mushra for proposed model}
\label{sec:mushra_proposed}
This MUSHRA test asked 40 listeners to compare 3 versions of each of 74 segments.
The upper anchor is an original recording, compared to the same segment synthesised by \modelname.
The lower anchor is WGANSing, representative of SOTA, described in \refsec{sec:baseline}.
Note that the MUSHRA methodology requires that we train the models compared on the same data, in order to remove dataset bias.
Because of this, WGANSing was trained on the same corpus we used for the proposed system.


\section{Results}
\label{sec:results:general}
\begin{figure}
\centering
\includegraphics[width=\textwidth / 2]{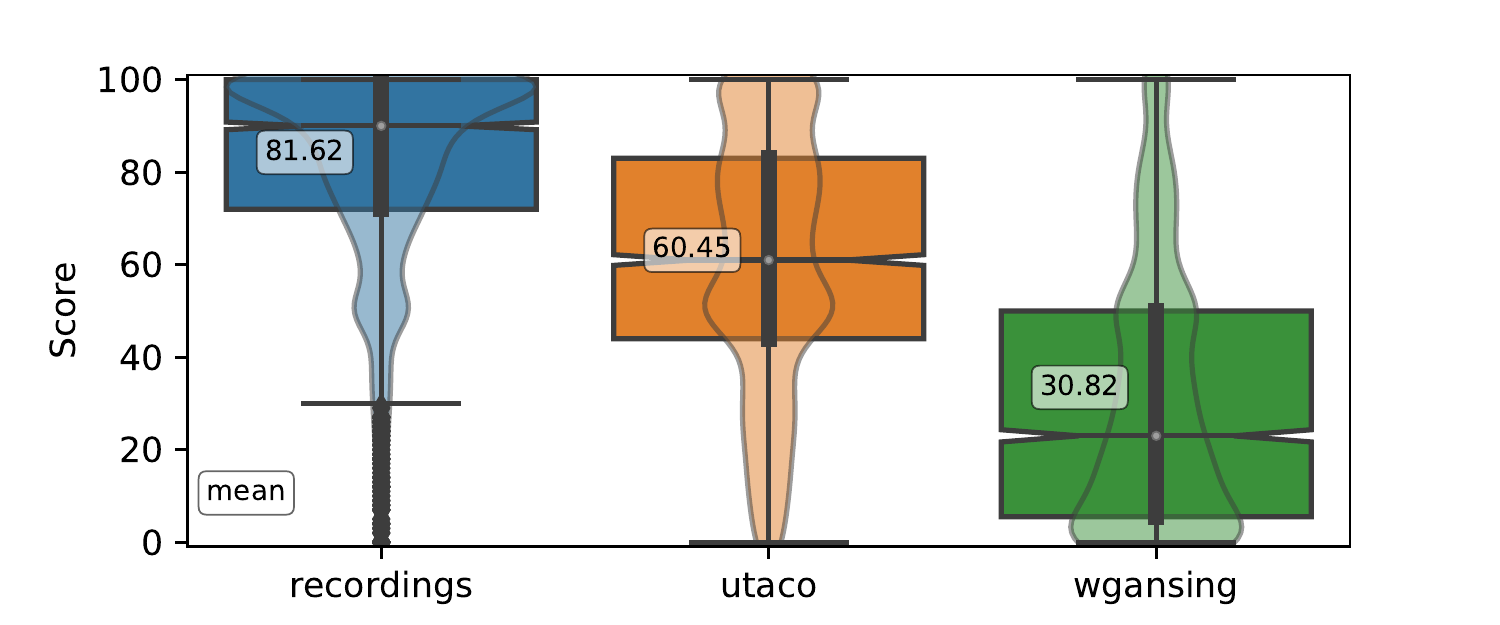}
\caption{Box and violin plot of MUSHRA for \modelname, trained on our corpus.}
\label{fig:mushra_boxplot}
\end{figure}

The MUSHRA test results, presented in \reffigure{fig:mushra_boxplot}
show that the mean relative MUSHRA score for \modelname\ is 60.45, compared to recordings which have 81.62.
Compared to the baseline, at 30.82, there is a significant improvement in naturalness over the SOTA.
Upon closer inspection, we observed that most of \modelname's segments in the lower score quartile (whose scores range between 26-50) contain either a vocoder glitch (there are 3 in the mushra dataset) or mumbled/mispronounced words.
Segments in the upper two quartiles, which obtain average scores in the range 61-77,
seem to show much less of these problems\footnote{We will release the samples if the paper is approved for publishing.}.
The main improvement might be due to the clarity and expressivity of our model compared to the baseline model.
The MUSHRA listeners commented on the sound quality of the baseline defining it \qt{muffled}, \qt{glitchy}, \qt{poor}.
Regarding the naturalness of the voice, the baseline was perceived as \qt{monotonous} and \qt{too regular to sound natural}, although at inference time it was fed oracle F0 and phoneme timings from the original recordings.

In contrast to \cite{Blaauw2017}, \modelname\ sings in tune autonomously.
We found that training on non-autotuned data produces a model that often goes out of tune.
\modelname\ performs best on simpler songs, which do not include very high or low-pitched notes, or phonemes sustained for a long time. Thich was expected given these contexts are under-represented in the data (see \reffig{fig:pitch_shift}).
Our model suffers from some articulation difficulties on low-frequency phoneme combinations.
MUSHRA listeners noticed this, and commented about \qt{mispronunciations}, but this was expected since we used only $\sim$2 hours of unaugmented data.
Since the dataset did not contain vibrato annotations, no explicit indication of it was included in the note embeddings.
Nevertheless, we observed that the model learns to reproduce a good vibrato, and apply it in the right places -- on longer sustained notes -- according to the musical context.

The two main drawbacks of \modelname\ are due to the nature of the architecture of choice.
First, we verified that when a long rest, i.e. silence, is encountered in the score, the duration of the silence produced by the encoder is completely unpredictable.
We observed that the concentration of the attention weights decreases significantly on pause tokens, leading to instability.
This issue, similar to word-skipping, seems to be common to \periphrasis\ models in general \cite{Shen2018,Ping2017}, but is more severe in music, where pauses of any given length are frequent and essential to the musical context and rhythm.
Another possibly related drawback is that \modelname\ seems to produce notes that are slightly too long or too short, losing the rhythm over time.
This could be due to the combined effect of lack of data and augmentations.
Changing the bpm of a song affects all durations uniformly, so any discrepancies between score duration and phoneme duration in the training data are correlated among augmented versions of the same song, which would essentially lead to overfitting on errors.
In our architecture it is not possible to directly control the timing of the attention matrix.
None of the two problems reduce its ultimate ability to synthesise a singing voice. Two easy workarounds are cutting the scores on rests, and editing the tempo in post-processing in case the timing problem manifests itself.
We observed that low-scored samples from the MUSHRA test seem to suffer more frequently from vocoder glitches and mumbling than attention instabilities.
We attribute the former to the lack of data.

\section{Discussion}
\label{sec:discussion}
In this paper we presented \modelname, to the best of our knowledge, the first use of the attention mechanism to generate pitch and timing in the field of singing synthesis.
What sets the system apart from other techniques, other than the noticeable improvement in naturalness over the SOTA, is the complete lack of need for explicitly modelling many parts of the song synthesis process.
The \periphrasis\ architecture is capable of autonomously modelling F0 patterns, vibratos, and inserting vibrato in the right context.
Training \modelname\ requires a dataset whose size is the same order of magnitude as typical datasets in the field \cite{Gomez2018arXiv}.
Its only input is the musical score with lyrics to be synthesised: it requires no explicit modelling of any feature of a singing voice.
As noted earlier, our embeddings are leaner than those in \cite{Blaauw2017}, which also embed the previous/next phoneme, as well as other linguistic features and more importantly phoneme length.
We hypothesise that the LSTM in the encoder plus the attention makes them redundant.

Many of the previous singing synthesis systems require a separate model for allocating phoneme duration inside of a note.
To increase naturalness, a model is sometimes needed to emulate the small imperfections in timing found in natural singing, such as in \cite{Hono2018}.
In other words, many duration values needed for the synthesis are not unambiguously specified by the score.
Attention dispenses with the need to model duration, and is conceptually simpler than previous systems proposed in the literature.
Therefore, one major improvement is the reduced amount of modelling work needed to create a singing synthesis system with the method we described here.
However, the main drawbacks of employing an \periphrasis\ architecture are in the precision of timing, and are also a direct result of the attention model used. Some previous work (e.g. \cite{Ping2017,Ren2019}) already tackled the same problems.
We propose that \modelname\ can be stabilised with further work, and that the benefits of a stable attention model would ultimately justify its use.

\periphrasis\ models enjoy much active research work, and all new extensions of such architectures, such as speaker identity, language conditioning and style conditioning can potentially be applied to \modelname\ immediately.

\bibliographystyle{IEEEtran}
\bibliography{bibliography}

\end{document}